\documentclass[reqno]{amsart}
\usepackage{amsfonts}
\usepackage{amsthm}
\usepackage{amssymb}
\usepackage{dsfont}
\usepackage{amscd}
\usepackage{color}
\usepackage{tikz}
 \usetikzlibrary{calc,decorations.markings,matrix,decorations.pathmorphing}

\theoremstyle{definition}

\theoremstyle{remark}

\numberwithin{equation}{section}

\def\ra{\rightarrow}
\def\iy{\infty}
\def\hf{{1\over 2}}

\def\be{\begin{equation}}
\def\ee{\end{equation}}
\def\ba{\begin{eqnarray*}}
\def\ea{\end{eqnarray*}}
\def\bae{\begin{eqnarray}}
\def\eae{\end{eqnarray}}
\def\bc{\begin{center}}
\def\ec{\end{center}}
\renewcommand{\ss}{\scriptstyle}

\begin{document}

\title{Equilibrium problems for Raney densities}
\author{Peter J. Forrester} \address{Department of Mathematics and Statistics, The University of Melbourne, Victoria 3010, Australia}
\author{Dang-Zheng Liu} \address{Wu Wen-Tsun Key Laboratory of Mathematics, USTC, CAS and School of Mathematical Sciences, University of Science and Technology of China, Hefei 230026, P. R. China}
\author{Paul Zinn-Justin} \address{Laboratoire de Physique Th\'eorique et Hautes \'Energies, CNRS UMR 7589 and
Universit\'e Pierre et Marie Curie (Paris 6), 4 place Jussieu, 75252 Paris cedex 05, France}

\begin{abstract}
The Raney numbers are a class of combinatorial numbers generalising the Fuss--Catalan numbers. They
are indexed by a pair of positive real numbers $(p,r)$ with $p>1$ and $0 < r \le p$, and form the moments
of a probability density function. For certain $(p,r)$ the latter has the interpretation as the density of squared
singular values for certain random matrix ensembles, and in this context equilibrium problems characterising
the Raney densities for $(p,r) = (\theta +1,1)$ and $(\theta/2+1,1/2)$ have recently been proposed. Using two different
techniques --- one based on the Wiener--Hopf method for the solution of integral equations and the other on an analysis of the algebraic equation satisfied by the Green's
function --- we establish the validity of the equilibrium problems for general $\theta > 0$ and similarly use both methods
to identify the equilibrium problem for $(p,r) = (\theta/q+1,1/q)$, $\theta > 0$ and $q \in \mathbb  Z^+$. The Wiener--Hopf method
is used to extend the latter to parameters $(p,r) = (\theta/q + 1, m+ 1/q)$ for $m$ a non-negative integer, and also to identify the
equilibrium problem for a family of densities
with  moments given by certain binomial coefficients.
\end{abstract}

\maketitle
\section{Introduction}
The Raney numbers
\be\label{1.1}
R_{p,r}(k) = \frac{r}{pk + r}\binom{pk + r}{k}, \quad k=0, 1, \dots
\ee
indexed by the pair $(p,r)$ with $p>1$ and $0<r\leq p$ naturally appear in combinational problems generalizing the ballot problem \cite{GKP89}. Recent studies \cite{PZ11,FL14,MNPZ14} have drawn attention to the fact that for particular $(p,r)$ the Raney numbers also occur in random matrix theory. In this setting $ \{{R_{p,r}(k)} \}_{k=0,1,\dots}$ corresponds to the moments of the (normalized) global density of the squared singular values for certain random matrix ensembles. An example, applying to the case $p=M+1, M\in \mathbb Z^+$, and $r=1$ is the product $X_1 \cdots X_M$ of $N\times N$ standard Gaussian matrices in the limit $N\ra \iy$, with the squared singular values $\lambda_j$ first scaled by $\lambda_j \mapsto \lambda_j N^M$ \cite{AGT10,BJLNS11,BBCC11,NS06,Ne14,NS14}.
We remark that the Raney numbers $(M+1,1)$ are better known as the Fuss--Catalan numbers, see
e.g.~\cite{PZ11}.

Let $\rho_{\ p,r}^{(1)} (x)$ denote the density with moments (\ref{1.1}). It is known \cite{Ml10} that there is an $L>0$ such that the support is $[0, L]$. With the resolvent, or Green's function, defined by
\be\label{1.2}
G_{p,r}(z) = \int_0^L \frac{\rho_{\ p,r}^{(1)} (x)}{z-x} dx,
\ee
it follows by expanding the denominator of the integrand using the geometric series that $G_{p,r}$ is in fact the generating function for Raney numbers,
$$
zG_{p,r}(z) = \sum_{k=0}^{\iy} R_{p,r}(k)z^{-k}, \quad |z| > L.
$$
One remarkable property of this generating function is that it satisfies the algebraic equation \cite{MPZ13}
\be\label{1.3}
(zG_{p,r})^{p/r} - z(zG_{p,r})^{1/r} + z = 0.
\ee

A feature of global densities in random matrix theory is that they admit characterisations as the solution of equilibrium problems (see e.g.~\cite{Fo10,PS11}). This motivated two of the present authors  to initiate a study into the specification of equilibrium problems for Raney densities,  and so supplement the characterisation (\ref{1.3}). Specifically, in \cite{FL14}, this program was carried out for parameters $(p, r) = (\theta+1, 1),\ \theta \ge 1$ i.e.~the Fuss--Catalan case, and $(p, r) = (\theta/2 + 1, 1/2),\ \theta>0$ with the latter only a conjecture beyond the cases $\theta = 1,2$. It is the purpose of the present paper to show that both of these equilibrium problems do indeed correspond to the appropriate Raney density for general $\theta>0$, and moreover to specify equilibrium problems for Raney densities $(p,r) = (\theta/q+1, 1/q),\ \theta>0$ and $q\in \mathbb Z^+$. Our tools are two methods introduced by the third named co-author of the present paper in the context of the study of equilibrium problems for certain matrix models \cite{ZJ00,ZJ00a}.

To be more specific in relation to these methods, we recall that by interpreting results from \cite{CR13} it was shown \cite[Cor.4.2]{FL14} that for $\theta>1$ at least, $\rho_{\ \theta+1,1}^{(1)}$ minimizes the energy functional
\begin{multline}\label{1.4}
E_{\theta+1,1}[\rho^{(1)}(y)] = \theta\int_0^L y^{1/\theta}\rho^{(1)}(y) dy - \hf \int_0^Ldy \int_0^Ldy'\rho^{(1)}(y)\rho^{(1)}(y') \\
\times \log{(|y^{1/\theta} - (y')^{1/\theta}| |y - y'|)}
\end{multline}
with
\begin{equation}\label{1.4a}
L = \theta(1 + 1/\theta)^{1+\theta}.
\end{equation}
Here and below the subscripts on the energy functional match the parameters of the corresponding moments.
Taking the functional derivative of (\ref{1.4}) with respect to $\rho^{(1)}$ characterises the latter as the solution of the integral equation
\be\label{1.5}
\int_0^L dy'\rho^{(1)}(y') \log{(|y^{1/\theta} - (y')^{1/\theta}| |y - y'|)} = \theta y^{1/\theta} + C, \quad 0<y<L.
\ee
The technique of \cite{ZJ00} is to seek a solution of (\ref{1.5}) using the Wiener--Hopf method. As demonstrated in \cite{ZJ00}, this method has the appealing feature of leading directly to the moments. In contrast, the technique of \cite{ZJ00a} provides a method for us to begin with the algebraic equation (\ref{1.3}) in the case $(p, r) = (\theta+1, 1), \theta \in \mathbb Z^+$ and deduce from this the integral equation (\ref{1.5}).
Both methods can also be used in relation to the case $(p, r) = (\theta/q+1, 1/q)$, $q \in \mathbb Z^+$. This is done in Sections
\ref{s3.1} and \ref{s3.2} respectively.

Using the Wiener--Hopf method we are also able to extend the latter case, and so identify the equilibrium problem for the Raney densities with parameters
$(p, r) = (\theta/q+1, m+1/q)$ for $ m \leq (\theta-1)/q + 1 $ an integer. This is done in Section \ref{s4.1}.
In Section \ref{s4.2}, we show that a slight modification of the Wiener--Hopf  analysis of Section \ref{s3.1} leads to the determination of
an equilibrium problem  for a family of densities for which the moments
are certain binomial coefficients.

In the opening paragraph, it was commented that the Fuss--Catalan case of the Raney numbers
for parameters $(p,r) = (M+1,1)$ relates to the squared singular values of the product of $M$
standard Gaussian matrices. It must therefore also be that the squared singular values of such
a random matrix product relate to the energy functional (\ref{1.4}). In the Appendix we will show how this result can be anticipated.

\section{The equilibrium problem for $(p, r) = (\theta+1, 1)$}
\subsection{Wiener--Hopf method}\label{s2.1}

We begin by  scaling the variables in (\ref{1.5}) so that the support is on $[0, 1]$. Differentiating with respect to $y$, and multiplying both sides by $y$ then gives
\be\label{2.1}
{\rm PV} \int_0^1 \frac{\rho^{(1)}(y')}{1-y'/y}dy' + {\rm PV} \frac{1}{\theta} \int_0^1\frac{\rho^{(1)}(y')}{1-(y'/y)^{1/\theta}} \, dy' = L^{1/\theta - 1} y^{1/\theta}, \quad 0<y<1,
\ee
where PV denotes the principal value.
We now substitute $y = e^{-t}, y' = e^{-s},\ 0\leq t<\iy,\ 0\leq s<\iy$, and set $e^{-s}\rho^{(1)}(e^{-s}) = \phi(s)$. The integral equation then reads
\be\label{2.2}
{\rm PV} \int_{0}^\iy \Big (\frac{1}{1-e^{-(s-t)}} + \frac{1}{\theta} \frac{1}{1-e^{-(s-t)/\theta}} \Big )\phi(s) \,ds =  L^{1/\theta - 1} e^{-t/\theta}, \quad 0<t<\iy.
\ee

Let us extend the range of validity of (\ref{2.2}) by defining the RHS to equal $R(t)$ where
\be\label{2.2a}
R(t) = \left \{ \begin{array}{ll}
L^{1/\theta - 1} e^{-t/\theta}, & 0 \leq t < \iy \\
r(t), & -\iy < t < 0. \end{array} \right.
\ee
The exact functional form of $r(t)$ is unknown, but inspection of the LHS of (\ref{2.2}) shows that it is bounded.
Since the kernel in (\ref{2.2}) is a function of difference variables, unlike the situation in (\ref{2.1}), we can use a Fourier transform to factorise the LHS. This can be carried out by multiplying both sides by $e^{itz}$ and integrating over $t$ from $-\iy$ to $\iy$ to give
\be\label{2.3}
{\rm PV} \bigg( \int_{-\iy}^{\iy} \frac{e^{itz}}{1-e^t} dt + \frac{1}{\theta}\int_{-\iy}^{\iy}\frac{e^{itz}}{1-e^{t/\theta}}dt \bigg) \int_0^{\iy}\phi(s)e^{isz}ds = \int_{-\iy}^\iy R(t)e^{itz \, }dt.
\ee

Using residue calculus we can check that
\begin{equation}\label{2.3a}
{\rm PV} \,  \int_{-\iy}^{\iy} \frac{e^{itz}}{1-e^t} \,dt = -\pi i \frac{\cosh{\pi z}}{\sinh{\pi z}}
\end{equation}
and thus
$$
{\rm PV} \,  \frac{1}{\theta}\int_{-\iy}^{\iy} \frac{e^{itz}}{1-e^{t/\theta}} \,dt = -\pi i \frac{\cosh{\pi z\theta}}{\sinh{\pi z\theta}}.
$$
With the first factor on the LHS of (\ref{2.3}) denoted $K(z)$ these results together with the use of a simple identity between hyperbolic functions shows
\begin{equation}\label{2.3b}
K(z) = -\pi i \frac{\sinh{\pi z(1+\theta)}}{\sinh{\pi z}\sinh{\pi z\theta}}.
\end{equation}

In keeping with the general strategy of the Wiener--Hopf method (see e.g.~\cite{LA07}) we want to factor $K(z)$ according to $K(z) = K_+(z)/K_-(z)$ where $K_+(z)$ is analytic for ${\rm Im}(z)>-\epsilon \ (\epsilon>0), K_-(z)$ is analytic for ${\rm Im}(z)<0$ and both $K_+(z)/z$ and $K_-(z)/z$ are bounded in the corresponding one of these half planes. For this purpose, use of the functional equation for the gamma function in the form
\begin{equation}\label{2.3c}
\Gamma(iz)\Gamma(1-iz) = \frac{\pi}{i\sinh{\pi z}}
\end{equation}
allows us to write
$$
K(z) = \frac{\Gamma(iz)\Gamma(1-iz)\Gamma(iz\theta)\Gamma(1-iz\theta)}{\Gamma(iz(1+\theta))\Gamma(1-iz(1+\theta))}.
$$

Consider now the factorisation
\begin{equation}\label{2.5}
K_+(z) = \frac{\Gamma(1-iz)\Gamma(1-iz\theta)}{\Gamma(1-iz(1+\theta))}e^{icz}, \qquad
\frac{1}{K_-(z)} = \frac{\Gamma(iz)\Gamma(iz\theta)}{\Gamma(iz(1+\theta))}e^{-icz},
\end{equation}
where $c$ is yet to be determined. We observe that $K_+(z)$ is analytic for ${\rm Im}(z)>-{\rm min}(1, 1/\theta)$ while $K_-(z)$ is analytic for ${\rm Im}(z)<0$. The value of $c$ is to be chosen so that $K_{\pm}(z)/z$ are bounded in their respective half planes. Use of Stirling's formula shows that this is achieved by choosing
\be\label{2.6x}
c = -\big( (1+\theta)\log(1+\theta) - \theta\log(\theta) \big),
\ee
and we then have for large $|z|$
\be\label{2.7i}
{1 \over z} K_-(z) \sim \frac{1}{\sqrt{2\pi}} \sqrt{\theta \over 1+\theta} \sqrt{i \over z}, \quad  {1 \over z} K_+(z) \sim \sqrt{2\pi}\sqrt{\theta  \over 1+\theta} \sqrt{-i \over z}.
\ee

From the definition of $\phi(s)$ in terms of $\rho^{(1)}$ below (\ref{2.1}), we have that the Fourier transform $\int_0^\iy \phi(s)e^{isz}ds$ is analytic for ${\rm Im}(z) \geq -\epsilon$ for some $\epsilon >0$. Also $\int_{-\iy}^0 r(t)e^{itz}dt$ is analytic for ${\rm Im}(z)<0$. Replacing the first factor in (\ref{2.3}) by $K_+(z)/K_-(z)$, multiplying through by $K_-(z)$ and simplifying the Fourier transform of $R(t)$ by substituting (\ref{2.2a}) and evaluating the integral in the range $0 \leq t < \iy$ shows that (\ref{2.3}) can be rewritten to read
\be\label{2.6}
K_+(z) \int_0^\iy \phi(s)e^{isz}ds = K_-(z)\Big( \int_{-\iy}^0 r(t)e^{itz} \,dt + L^{1/\theta - 1}\frac{\theta}{1-i\theta z} \Big).
\ee

The crucial feature of (\ref{2.6}) from the viewpoint of the Wiener--Hopf method is that the LHS is analytic and bounded in the half plane ${\rm Im}(z) \geq -\epsilon$ for some $\epsilon>0$, while the RHS is analytic and bounded in the half plane ${\rm Im}(z)<0$, except for a simple pole at $z=-i/\theta$. Thus the regions of analyticity overlap, so the RHS of (\ref{2.6}) is an analytic continuation of the LHS into the region ${\rm Im}(z)<0$. Hence the analytic function corresponding to the LHS of (\ref{2.6}) is analytic except for a simple pole at $z=-i/\theta$ and furthermore decays for $|z| \ra \iy$. The only possibility is that this function is proportional to $1/(z+i/\theta)$ and thus
\be\label{2.7}
K_+(z) \int_0^\iy \phi(s)e^{isz}ds = \frac{A}{z+i/\theta}.
\ee
The value of $A$ must equal the residue at $z=-i/\theta$ on the RHS of (\ref{2.6}), telling us that
\be\label{2.7a}
A = -{L^{1/\theta - 1} \over i} K_-\Big(-\frac{i}{\theta}\Big) = {i \over \theta} L^{1/\theta - 1} e^{c/\theta} = {i \over L \theta},
\ee
where to obtain the last equality use has been made of (\ref{1.4a}) and (\ref{2.6x}).
Hence we have an explicit formula for the general (complex) moments of $\rho^{(1)}$,
\begin{align}\label{2.8}
\int_0^\iy \phi(s)e^{isz}ds
& = \int_0^1 \rho^{(1)}(x)x^{-iz} \, dx
\nonumber
\\ & = \frac{1}{L}\frac{e^{-icz}}{\theta z/i+1} \frac{\Gamma(1-iz(1+\theta))}{\Gamma(1-iz)\Gamma(1-iz\theta)}.
\end{align}

Recalling that $c$ is given by (\ref{2.6}), and
with $L$ given by (\ref{1.4a}), setting $z=in, n\in \mathbb N_0$ in (\ref{2.8}) gives for the integer moments
\be\label{2.9}
L^{n+1}\int_0^1\rho^{(1)}(x)x^n \, dx = \frac{1}{\theta n+1} \binom{n(1+\theta)}{n}.
\ee
Simple manipulation of this expression reveals that it is precisely the Raney number $R_{\theta+1, 1}(n)$ as specified by (\ref{1.1}). Hence, by use of the Wiener--Hopf method, we have shown that the solution of the integral equation (\ref{1.5}) corresponding to the minimization of the energy functional (\ref{1.4}) is given by the Raney density $\rho^{(1)}_{\theta+1, 1}$ for general $\theta>0$.

\subsection{Analysis of the algebraic equation}\label{s2.2}

For simplicity of presentation, we will assume throughout this subsection that $\theta$ is a positive integer. Writing $w(z)=zG_{\theta+1, 1}(z)$ the algebraic equation (\ref{1.3}) reads
\be\label{2.10}
w^{\theta+1} - zw + z = 0.
\ee
Our aim is to show that this implies the integral equation (\ref{2.1}).

The equation (\ref{2.10}) defines a curve in $\mathbb{C}^2$ (plus two points at infinity). By analysing the differential of the equation we find that the curve is smooth and has a branch cut at $(0, 0)$ with the analytic behaviour $w \sim (-z)^{1/(\theta+1)}$ as $z\ra0$, as well as a square root branch cut at $w_c = 1/\theta+1, z_c=\theta^{-\theta}(1+\theta)^{1+\theta}$. As $z\ra\iy$, either $w$ stays finite as does $-zw+z$ and thus $w\ra1$, or $w\ra\iy$ with the asymptotic behaviour $w \sim z^{1/\theta}$.

\begin{figure}
\begin{tikzpicture}
\def\p{3}
\foreach\i in {0,...,\p}
{
\draw (0,-\i) -- ++(3,0) -- ++ (1,0.8) -- ++(-3,0) -- cycle;
\node[fill,circle,inner sep=1pt,label={[yshift=-2mm,xshift=1mm]left:$\ss 0$}] (z\i) at (2,-\i+0.4) {}; 
}
\node at (5,0.4) {$w(z)$};
\node at (5,-0.6) {$w^\star_1(z)$};
\node at (5,-1.6) {$\vdots$};
\node at (5,-2.6) {$w^\star_{p-1}(z)$};

\node[fill,circle,inner sep=1pt,label=right:$\ss z_c$] (c0) at (2.8,0.4) {};
\node[fill,circle,inner sep=1pt,label=right:$\ss z_c$] (c1) at (2.8,-0.6) {};
\foreach\i in {1,...,\p}
{
\ifnum\i>1
\draw[fill=lightgray] (z\i.center) ++(-1.5,0) -- ++(0,0.6) -- ++(1.5,0) -- ++(0,-0.6);
\draw[dotted] (z\i) ++(0,0.6) -- ++(0,0.4);
\draw[dotted] (z\i) ++(-1.5,0) ++(0,0.6) -- ++(0,0.4);
\fi
\draw[decorate,decoration={zigzag,segment length=0.12cm,amplitude=1pt}] (z\i) -- ++(-1.5,0);
}
\draw[fill=lightgray] (z1.center) -- ++(0,0.6) -- ++(0.8,0) -- ++(0,-0.6);
\draw[dotted] (z1) ++(0,0.6) -- ++(0,0.4);
\draw[dotted] (c1) ++(0,0.6) -- ++(0,0.4);
\draw[decorate,decoration={zigzag,segment length=0.12cm,amplitude=1pt}] (z0) -- (c0) (z1) -- (c1);
\end{tikzpicture}
\caption{Analytic structure of $w(z)$.}\label{F1}
\end{figure}
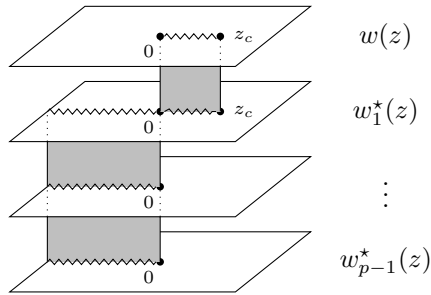

The solutions of (\ref{2.10}), totalling $\theta+1$ in number (this is why we restrict attention to $\theta$ a positive integer) can be considered as a Riemann surface corresponding to copies of the complex $z$-plane joined along appropriate branch cuts; see Figure \ref{F1}. From the definition of $w(z)$ in terms of the resolvent we require $w(z) \ra 1$ as $z \ra \iy$. This determines the physical sheet, with the other solutions $w_i^*(z) \ (i=1, \ldots, \theta)$ say, obtained by going through the physical cut. The monodromy transformations are the elementary transposition $(w, w_1^*)$ and the $\theta$-cycle $(w_1^*, \ldots, w_\theta^*)$.

This viewpoint suggests performing the change of variables $z=x^\theta$ in (\ref{2.10}). On the one hand this collapses all the sheets $w_i^*(z)$ into one single sheet $w^*(x)$, while on the other hand the physical sheet $w(z)$ now has multiple copies of the physical cut (see Figure \ref{F2}). The important point is that $w^*(x)$ is analytic outside the physical cut $[0, x_c]$ and its discontinuity there is the same as that of $w(z)$ up to a sign and to reparametrisation $z \ra x$. From the definition of $w(z)$ in terms of the Green's function the discontinuity is equal to $2\pi iz\rho_{\theta+1, 1}^{(1)}(z)$.

\begin{figure}
\begin{tikzpicture}
\foreach\i in {0,...,1}
{
\draw (0,-\i) -- ++(3,0) -- ++ (1,0.8) -- ++(-3,0) -- cycle;
\node[fill,circle,inner sep=1pt,label={[yshift=-2mm,xshift=1mm]left:$\ss 0$}] (z\i) at (2,-\i+0.4) {}; 
}
\node at (5,0.4) {$w(x)$};
\node at (5,-0.6) {$w^\star(x)$};
\node[fill,circle,inner sep=1pt,label=right:$\ss x_c$] (c0) at (2.8,0.4) {};
\node[fill,circle,inner sep=1pt,label=right:$\ss x_c$] (c1) at (2.8,-0.6) {};
\draw[fill=lightgray] (z1.center) -- ++(0,0.6) -- ++(0.8,0) -- ++(0,-0.6);
\draw[dotted] (z1) ++(0,0.6) -- ++(0,0.4);
\draw[dotted] (c1) ++(0,0.6) -- ++(0,0.4);
\draw[decorate,decoration={zigzag,segment length=0.12cm,amplitude=1pt}] (z0) -- (c0) (z1) -- (c1) (z0) -- ++(-0.3,0.3) node[fill,circle,inner sep=1pt] (z0b) {} (z0) -- ++(-0.6,-0.3) node[fill,circle,inner sep=1pt] (z0c) {};
\draw[dotted] (z0) -- ++(0,0.4);
\draw[dotted] (z0b) -- ++(0,0.4);
\draw[dotted] (z0c) -- ++(0,0.4);
\end{tikzpicture}
\caption{Analytic structure of $w(x)$.}\label{F2}
\end{figure}

We now introduce
\be\label{G}
\tilde{G}_{\theta+1,1}(x) = \int_0^{L^{1/\theta}}\frac{(x')^\theta \rho_{\theta+1,1}^{(1)}((x')^\theta)}{x-x'} \,dx'.
\ee
This has discontinuity $2\pi iz\rho_{\theta+1,1}^{(1)}(z), z=x^\theta$, along its cut, thus cancelling out the discontinuity of $w^*(x)$. Hence $w^*(x) + \tilde{G}(x)$ is an entire function. From (\ref{2.10}) with $z=x^\theta$ we can check that for $x\ra\iy, w^*(x)=x-1/\theta+o(1)$, while it follows from (\ref{G}) that $\tilde{G}_{\theta+1, 1}(x) \sim O(1/x)$. Consequently, by Liouville's theorem
\begin{align}\label{G1}
w^*(x) & = x - \tilde{G}_{\theta+1, 1}(x) - \frac{1}{\theta} \\
& = x - {x \over \theta} \int_0^L\frac{\rho_{\theta+1,1}^{(1)}(z')}{x-(z')^{1/\theta}} \, dz',
\end{align}
where the second line follows by a change of variables in (\ref{G}), and simple manipulation.

Next consider (\ref{G}) averaged over the cases that x approaches the physical cut from below. Noting from the definition of $w^*(x)$ that
$$
w^*(x+i0) + w^*(x-i0) = w(x-i\theta) + w(x+i\theta)
$$
this procedure gives
\be\label{G1a}
\hf\big( w(x+i0) + w(x-i0) + \tilde{G}_{\theta+1, 1}(x+i0) + \tilde{G}_{\theta+1, 1}(x-i0) \big) = x - {1 \over \theta}.
\ee
In the variable $z=x^\theta$ it follows from this that
$$
G_{\theta+1, 1}(z \pm i0) + \frac{1}{z} \tilde{G}_{\theta+1, 1}(z^{1/\theta} \pm i0) = z^{1/\theta-1} - \frac{1}{\theta z}.
$$
Finally, we substitute the definitions of $G_{\theta+1, 1}$ and $\tilde{G}_{\theta+1, 1}$ and perform a simple manipulation to obtain
$$
\int_0^L \frac{\rho_{\theta+1,1}^{(1)}(z')}{z - z'} \, dz' + \frac{z^{1/\theta-1}}{\theta} \int_0^L \frac{\rho(z')}{z^{1/\theta}-(z')^{1/\theta}} \, dz' = z^{1/\theta-1},
$$
which we recognise as equivalent to (\ref{2.1}).

\section{The equilibrium problem for $(p, r) = (\theta/q+1, 1/q)$}
\subsection{Wiener--Hopf method}\label{s3.1}

In \cite{FL14} it was proposed that $\rho_{\theta/2+1, 1/2}^{(1)}$ minimizes the energy functional
\begin{multline}\label{3.1}
E_{\theta/2+1, 1/2} [\rho^{(1)}(y)] = \theta\int_0^L y^{1/\theta} \rho^{(1)}(y)dy - \hf\int_0^L dy \int_0^Ldy'\rho^{(1)}(y)\rho^{(1)}(y') \\
\times \log{\bigg( \frac{|y^{1/\theta}-(y')^{1/\theta}|}{|y^{1/\theta}+(y')^{1/\theta}|}|y-y'| \bigg)},
\end{multline}
which could be verified for $\theta=1,2$. We observe that with $q \in \mathbb Z^+$, (\ref{1.4}) and (\ref{3.1}) are the cases $q=1$ and $q=2$ respectively of the energy functional
\begin{multline}\label{3.2}
E_{\theta/q+1, 1/q} [\rho^{(1)}(y)] = \theta\int_0^L y^{1/\theta} \rho^{(1)}(y)dy - \hf\int_0^L dy \int_0^Ldy'\rho^{(1)}(y)\rho^{(1)}(y') \\
\times \log{\bigg( |y-y'| \prod_{p=0}^{q-1}|(y^{1/\theta} - \omega^p(y')^{1/\theta})^{\omega^p}| \bigg)},
\end{multline}
where $L = (\theta/q)(1 + q/\theta)^{1+\theta/q}$ and $\omega = e^{2\pi i/q}$. Our aim in this subsection is to use the Wiener--Hopf method to show that $\rho_{\theta/q+1, 1/q}^{(1)}$ does indeed minimise (\ref{3.2}).

Taking the functional derivative of (\ref{3.2}) with respect to $\rho^{(1)}(y)$, then differentiating with respect to $y$ shows that our
task is to solve the integral equation
\begin{multline}\label{3.3}
{\rm PV} \int_0^1 {\rho^{(1)}(y') \over y - y'} \, dy' + {1 \over \theta} y^{1/\theta - 1} \sum_{p=0}^{q-1} \omega^p \,
{\rm PV} \int_0^1  {\rho^{(1)}(y') \over y^{1/\theta} - \omega^p (y')^{1/\theta}} \, dy' \\
= L^{1/\theta - 1} y^{1/\theta - 1}, \qquad 0 < y < 1,
\end{multline}
where we have scaled the variables so the density is supported on $[0,1]$. Use of the simple summation formula
$$
\sum_{p=0}^{q-1} {\omega^p \over a - \omega^p b} = {q b^{q-1} \over a^q - b^q},
$$
and after multiplying both sides by $y$, this reads
\begin{equation}\label{3.3a}
{\rm PV}  \int_0^1 {\rho^{(1)}(y') \over 1 - y'/y} \, dy' + {q \over \theta} \, {\rm PV}  \int_0^1 {(y'/y)^{(q-1)/\theta} \rho^{(1)}(y') \over 1 - (y'/y)^{q/\theta}} \, dy'  =
L^{1/\theta - 1}  y^{1/\theta}, \: \: 0 < y < 1.
\end{equation}

We now make the same substitutions as in going from (\ref{2.1}) to (\ref{2.2}) to obtain
$$
{\rm PV} \int_{0}^\iy \Big (\frac{1}{1-e^{-(s-t)}} + \frac{q}{\theta} \frac{e^{-(q-1)(s-t)/\theta}}{1-e^{-q(s-t)/\theta}} \Big )\phi(s) \,ds = L^{1/\theta - 1}e^{-t/\theta}, \quad 0<t<\iy.
$$
Introducing $R(t)$ as in (\ref{2.2a}), where $r(t)$ is unknown but bounded, we can introduce Fourier transforms as in (\ref{2.3}) to obtain
\be\label{3.4}
{\rm PV} \bigg( \int_{-\iy}^{\iy} \frac{e^{itz}}{1-e^t} dt + \frac{q}{\theta}\int_{-\iy}^{\iy}\frac{e^{(q-1)t/\theta}e^{itz}}{1-e^{qt/\theta}} \,dt \bigg) \int_0^{\iy}\phi(s)e^{isz}ds = \int_{-\iy}^\iy R(t)e^{itz} \,dt.
\ee
The first term in the first factor on the LHS is evaluated according to (\ref{2.3a}), whereas for the second term simple manipulation of (\ref{2.3a}) shows
$$
{\rm PV} \, {q \over \theta} \int_{-\infty}^\infty {e^{(q-1) t/\theta} e^{itz} \over 1-e^{qt/\theta}} \, dt = - \pi i
{\cosh \pi (z \theta/q - i (q-1)/q) \over \sinh \pi (z \theta/q - i (q-1)/q)}.
$$
Denoting the first factor in (\ref{3.4}) by $K(z)$,  we see upon use of a simple hyperbolic function identity that
$$
K(z) = -\pi i \frac{\sinh{\pi (z(1+\theta/q)}-i(q-1)/q)}{\sinh{\pi z}\sinh{\pi (z\theta/q - i (q-1)/q)}}.
$$
We note that this reduces to (\ref{2.3b}) when $q=1$.

Use of the gamma function identity (\ref{2.3c}) allows us to write $K(z) = K_+(z)/ K_-(z)$ where
\begin{equation}\label{3.5}
K_+(z) = \frac{\Gamma(1-iz)\Gamma(1/q-iz\theta/q)}{\Gamma(1/q-iz(1+\theta/q))}e^{icz},
\quad
 \frac{1}{K_-(z)} = \frac{\Gamma(iz)\Gamma(iz\theta/q + (q-1)/q)}{\Gamma(iz(1+\theta/q) + (q-1)/q))}e^{-icz}.
\end{equation}
These, with  $c = -\big( (1+\theta/q)\log(1+\theta/q) - (\theta/q)\log(\theta/q) \big)$, have the same analytic properties as the case $q=1$ discussed in \S \ref{s2.1}.
Arguing as in the derivation of (\ref{2.7}), we see that the latter again holds, but with $K_+(z)$ as in (\ref{3.5}), and $A$ again
given by (\ref{2.7a}). Consequently the analogue of (\ref{2.8}) reads
$$
\int_0^1 \rho^{(1)}(x) x^{-i z} \, dx = {1 \over L}
{e^{-i c z} \over \theta z / i + 1} {\Gamma(1/q - i z(1 + \theta/q)) \over \Gamma(1 - i z) \Gamma(1/q - i z \theta/q)},
$$
and thus
\begin{align}\label{3.5a}
L^{n+1} \int_0^1 \rho^{(1)}(x) x^n \, dx & =
{1 \over \theta n + 1}
 \binom{n(1 + \theta/q) + 1/q - 1}{n} \nonumber \\
& ={1/q \over n (1 + \theta/q) + 1/q}
\binom{n(1 + \theta/q) + 1/q}{n},
\end{align}
which is precisely the Raney number $R_{\theta/q + 1, 1/q}(n)$ as specified by (\ref{1.1}).

\subsection{Analysis of the algebraic equation}\label{s3.2}
Here we restrict attention to the case that both $\theta$ and $q$ are positive integers. With $w(z) = z G_{\theta/q + 1,1/q}(z)$
the algebraic equation (\ref{1.3}) reads
\begin{equation}\label{3.6}
w^{\theta + q} - z w^q + z = 0.
\end{equation}
We want to show that this implies the integral equation (\ref{3.3}).

The fact that (\ref{3.6}) is formally identical to (\ref{2.10}) with $w \mapsto  w^q$, $ \theta \mapsto \theta/q$ suggests that
we begin by making the change of variables $z = u^{\theta/q}$ in keeping with that of \S \ref{s2.2}. The $w^*(u)$ has the
physical cut and a square root cut. We next change variables $u = x^q$. Introducing
$$
\tilde{G}_{\theta/q+1,1/q}(x) =
\int_0^{L^{1/\theta}}  { (x')^\theta \rho_{\theta/q + 1,1}^{(1)}((x')^\theta) \over x - x'} \, dx'
$$
and repeating the reasoning below (\ref{G}) we see that $w^*(x) + \sum_{l=0}^{q-1}
\omega^l \tilde{G}_{\theta/q+1,1/q}(
\omega^l x)$ is entire, while for $x$  large $w^*(x) = x + O(1/x)$ and thus by Liouville's theorem
$$
 w^*(x) + \sum_{l=0}^{q-1}
\omega^l \tilde{G}_{\theta/q+1,1/q}(
\omega^l x) = x.
$$
After taking averages as in (\ref{G1a}), manipulating using the fact that $\sum_{l=0}^{q-1} \omega^l = 0$
$(q=2,3,\dots)$, and reinstating the variable $z$ we see that (\ref{3.3}) results.

\section{The equilibrium problem for $(p,r) = (\theta/q + 1, m + 1/q)$ and a variant}
\subsection{The case $(p,r) = (\theta/q + 1, m + 1/q)$}\label{s4.1}
In \cite{FL14} it was proposed that the equilibrium problem for $(p,r) = (\theta + 1, 1)$ could be extended to that for
$(p,r) = (\theta+1, m+1)$ with $m=1,2,\dots$ by generalising the one body potential in (\ref{1.4}) from
$\theta y^{1/\theta}$ to $\sum_{l=0}^m c_l y^{(1 + l)/\theta}$ for suitable $c_l$. Here we will use the Wiener--Hopf method to verify this, and
moreover to show that the replacement of $\theta y^{1/\theta}$ in (\ref{3.2}) by
\begin{equation}\label{4.1x}
\sum_{l=0}^m c_{l,q}^{(m)} y^{(1 + l q)/\theta}
\end{equation}
for suitable $\{ c_{l,q} \}$ generalises the equilibrium problem for $(\theta/q+1,1/q)$ to that for $(\theta/q+1,m+1/q)$. In fact we will work
directly in this more general setting. Recalling then the derivation of (\ref{3.3a}), our task is to show that the moments of the solution of the
integral equation
\begin{multline}
{\rm PV} \int_0^1 {\rho^{(1)}(y') \over 1 - y'/y} \, dy' + {q \over \theta} \, {\rm PV}  \int_0^1 {(y'/y)^{(q-1)/\theta} \rho^{(1)}(y') \over 1 - (y'/y)^{q/\theta}} \, dy' \\
= {1 \over L} \sum_{l=0}^m {(1 + l q) c_{l,q}^{(m)}  \over \theta}
(L y)^{(1 + l q)/\theta},\label{4.1}
\end{multline}
where $0< y < 1$ are for a suitable choice of $\{ c_{l,q}^{(m)}  \}$ given by the Raney numbers with parameters $(p,r) = (\theta/q + 1, m+ 1/q)$.

In (\ref{4.1}) we introduce exponential variables as in going from (\ref{2.1}) to (\ref{2.2}). Then (\ref{3.4}) holds,
but with
$$
R(t) = \left \{
\begin{array}{ll} \displaystyle {1 \over L} \sum_{l=0}^m
{(1 + l q) c_{l,q}^{(m)}  \over \theta} L^{(1 + lq)/\theta} e^{- t (1 + lq)/\theta}, & 0 \le t < \infty \\
r(t), & - \infty < t < 0, \end{array} \right.
$$
where $r(t)$ is unknown but bounded. With $K_\pm(z)$ defined as in (\ref{3.5}), we thus have that the analogue of (\ref{2.6}) reads
\begin{equation}\label{4.2}
K_+(z) \int_0^\infty \phi(s) e^{i s z} \, ds =
K_-(z) \Big (
\int_{-\infty}^0 r(t) e^{it z} \, dt + {1 \over L}
\sum_{l=0}^m {c_{l,q}^{(m)} L^{(1 + lq)/\theta} \over 1 - i \theta z/(1 + lq)} \Big ).
\end{equation}
Arguing as in the derivation of (\ref{2.7}) we have that the analytic continuation of the RHS of (\ref{4.2}) must be given by
\begin{equation}\label{4.3}
\sum_{l=0}^m {\alpha_l \over 1 - i \theta z/(1 + l q)}, \qquad \alpha_l = { c_{l,q}^{(m)}  L^{(1 + lq)/\theta} \over L }
K_- \Big ( - {i (1 + lq) \over \theta} \Big ).
\end{equation}

Suppose we choose $\{c_{l,q} \}$ so that
\begin{equation}\label{4.4}
\alpha_l = {1 \over L}
 { \prod_{u=0}^{m-1}(1 - i (\theta + q) z/ (qu+1)) \over
\prod_{u=0,  u \ne l}^m (1 - i \theta z/(qu+1)} \bigg |_{z = - i (1 + ql)/\theta}.
\end{equation}
Then
\begin{equation}\label{4.5}
\sum_{l=0}^m {\alpha_l \over 1 - i \theta z/(1 + lq)} = {1 \over L}{   \prod_{u=0}^{m-1} (1 - i (\theta + q) z/ (qu+1)) \over
\prod_{u=0 }^m (1 - i \theta z/(qu+1)} =: Q(z),
\end{equation}
and we have
$$
\int_0^1 \rho^{(1)}(x) x^{-i z} \, dx = {1 \over L} {Q(z) \over K_+(z)}.
$$
Setting $z = in$, we make use of the recurrence for the gamma function to write the RHS in terms of a
binomial coefficient, then we make use of the identity between binomial coefficients implied by (\ref{3.5a}) to
conclude
\begin{equation}\label{4.6}
L^{n+1} \int_0^1 \rho^{(1)}(x) x^n \, dx = R_{\theta/q + 1, m + 1/q}(n),
\end{equation}
which is the required result.

It is of interest to give the explicit value of $\{c_{l,q}^{(m)} \}$ in (\ref{4.1x}). Actually, we know from \eqref{4.3} that 
\begin{equation}   c_{l,q}^{(m)}  L^{(1 + lq)/\theta-1} =\alpha_l e^{-c(1+lq)/\theta}\frac{\Gamma((1+lq)/\theta)\Gamma(l+1)}{\Gamma(l+1+(1+lq)/\theta) }
\nonumber
\end{equation}
(cf.~(\ref{3.5})), and from \eqref{4.4} that
\begin{equation}    L\alpha_l = \frac{1+mq}{1+lq} \frac{(-1)^l}{l!(m-l)!} \frac{\Gamma(m-l-(1+lq)/\theta)}{\Gamma(-l-(1+lq)/\theta) }.
\nonumber
\end{equation}
Combining  them together,  using the functional equation for the gamma function  twice (cf.~\eqref{2.3c}) and noting $c=-\log L$ we thus obtain 
\begin{equation}   c_{l,q}^{(m)}    =  \frac{1+mq}{1+lq} \frac{(-1)^{m-l}}{(m-l)!} \frac{\Gamma((1+lq)/\theta)}{\Gamma(1+l-m+(1+lq)/\theta) }.
\end{equation}
In the simplest case $m=1$, 
$$
L \alpha_0 = - {1 + q \over \theta}, \qquad L \alpha_1 = 1 + {1 + q \over \theta},
$$
 and
$$
c_{0,q}^{(1)} = - (1 + q), \qquad c_{1,q}^{(1)}  = {\theta \over 1 + q}.
$$
Specialising further to $q=1$ this gives that the one body potential corresponding to $R_{\theta +1, 2}$ is
${\theta \over 2} (y^{1/\theta} - 2/\theta)^2$. In the case $\theta = 2$ this is consistent with findings reported in
\cite{CR13}.
\subsection{A related equilibrium problem}\label{s4.2}
In the recent work \cite{FW14} the global density $\rho^{{\rm J}, (1)}_\theta(x)$ for the so-called Jacobi Muttalib--Borodin ensemble \cite{Mu95,Bo98},
specified by the eigenvalue probability density function proportional to
\be\label{J1}
\prod_{j=1}^N x_j^a(1-x_j)^b \prod_{1 \le j < k \le N} (x_k - x_j)( x_k^{1/\theta} - x_j^{1/\theta}), \quad 0<x_j<1,
\ee
was computed. Specifically, it was shown that the corresponding Green's function defined by (\ref{1.2}) with $L=1$ and
$\rho_{\ p,r}^{(1)}(x)  \mapsto  \rho^{{\rm J}, (1)}_\theta(x)$ satisfies the equation
\begin{equation}\label{G8}
z ( z {G}^{\rm J}(z) - 1)( z {G}^{\rm J}(z) + 1/\theta)^\theta = (z {G}^{\rm J}(z))^{\theta + 1}.
\end{equation}
It was then deduced from this that the moments $m_n^{\rm J}$ of the global density are given in terms of a binomial
coefficient according to
\begin{equation}\label{mL}
 {m}_n^{\rm J} = A^{-n}  \binom{(1 + \theta)n }{ n}, \qquad A  := (1 + \theta)^{1 + \theta} \theta^{-\theta}.
\end{equation}

The moments in (\ref{mL}) correspond to the case $(p,r) = (\theta+1,0)$ of the family of moments
\begin{equation}\label{mLF}
 {m}_n^{\rm J} (p,r) = A^{-n}  \binom{pn + r}{ n}.
 \end{equation}
For $ - 1 < r \le p-1$ these moments have recently been shown to specify a probability density \cite{MP13}.
Here we will establish an equilibrium problem for the cases $(p,r) = (\theta/q+1, 1/q-1)$, where
$q \in Z^+$.

We can immediately read off from (\ref{J1}) that the equilibrium problem corresponding to $ \rho^{{\rm J}, (1)}_\theta(x)$ is
to minimise the energy functional
\be\label{1.4j}
E_{\theta}[\rho^{(1)}(y)] = - \hf \int_0^1 dy \int_0^Ldy'\rho^{(1)}(y)\rho^{(1)}(y')
 \log{(|y^{1/\theta} - (y')^{1/\theta}| |y - y'|)},
\ee
and thus that $\rho^{{\rm J}, (1)}_\theta(x)$ satisfies the integral equation
\be\label{1.5j}
\int_0^1 dy'\rho^{(1)}(y') \log{(|y^{1/\theta} - (y')^{1/\theta}| |y - y'|)} = C, \quad 0<y<1
\ee
(cf.~(\ref{1.5})). In keeping the analysis presented in \S \ref{s2.1}, we can readily use the Wiener--Hopf method to deduce that
the moments of the solution of (\ref{1.5j}) are given by (\ref{mL}). More generally, we can follow the working of \S \ref{s3.1}
to show that the density minimising
\begin{multline}\label{4.13}
E_{\theta/q+1, 1/q-1}^{\rm J} [\rho^{(1)}(y)] =  - \hf\int_0^1dy \int_0^1dy'\rho^{(1)}(y)\rho^{(1)}(y') \\
\times \log{\bigg( |y-y'| \prod_{p=0}^{q-1}|(y^{1/\theta} - \omega^p(y')^{1/\theta})^{\omega^p}| \bigg)},
\end{multline}
has moments given by (\ref{mLF}) with $(p,r) = (\theta/q+1, 1/q-1)$, as we will  now demonstrate.

Thus, using working from \S \ref{s4.1}, and with $K_\pm(z)$ defined as in (\ref{3.5}), we see that the analogue of (\ref{4.2}) reads
\begin{equation}\label{4.2a}
K_+(z) \int_0^\infty \phi(s) e^{i s z} \, ds =
K_-(z) \Big (
\int_{-\infty}^0 r(t) e^{it z} \, dt \Big ).
\end{equation}
The RHS is analytic for Im$(z) < 0$, and so the analytic continuation of the LHS must be a constant $C$. Its
value is $C=1$, as determined by setting $z=0$. Substituting $z = in$ then shows
\be\label{s1a}
\int_0^1 \rho^{(1)}(x) x^n \, dx = {1 \over K_+(in)} = m_n^{\rm J},
\ee
where $m_n^{\rm J}$ is given by (\ref{mLF}) with $(p,r) = (\theta/q+1,1/q-1)$, as required.

\section*{Appendix}
\setcounter{equation}{0}
\renewcommand{\theequation}{A.\arabic{equation}}
Consider the random matrix product $X_1 \cdots X_M$, where each $X_i$ is an $N \times N$
standard complex Gaussian matrix. We know from \cite{AKW13} that the distribution of the squared singular values is proportional to
\begin{equation}\label{A1}
\prod_{1 \le j < k \le N} (x_k - x_j) \det \Big [ G_{0,M}^{M,0} \Big (
{\underline{\qquad} \atop j-1, 0^{M-1}} \Big | x_k \Big ) \Big ]_{j,k=1,N},
\end{equation}
where $G_{0,M}^{M,0} $ denotes a particular Meijer $G$-function, and $0^{M-1}$ denotes
0 repeated $M - 1$ times. Here we will demonstrate how it is possible to anticipate that
(\ref{A1}) relates to the energy functional (\ref{1.4}) with $\theta = M + 1$.

We will make the hypothesis that the feature determining the equilibrium problem is the
functional form of the distribution (\ref{A1}) for large arguments. In this limit, we can use
the asymptotic form \cite[Section 5.7]{Lu69}
$$
 G_{0,M}^{M,0} \Big (
{\underline{\qquad} \atop j-1, 0^{M-1}} \Big | x \Big )
\mathop{\sim}\limits_{x \to \infty} {(2\pi)^{(M-1)/2} \over M^{1/2}} x^{(j-1/2)/M - 1/2}
e^{- M x^{1/M}}.
$$
Substituting this in (\ref{A1}) and using the Vandermonde determinant evaluation
(see e.g.~\cite[eq.~(1.173)]{Fo10}) we obtain, up to proportionality, the large distances form
\begin{equation}\label{A3}
\prod_{l=1}^N x_l^{-1/2 + 1/2M} e^{-M x_l^{1/M}}
\prod_{1 \le j < k \le N} (x_k - x_j) (x_k^{1/M} - x_j^{1/M}).
\end{equation}
Introducing the change of scale $x_l \mapsto N^M x_l$, exponentiating the product of differences,
equating terms of order $N$ and taking a mean field viewpoint where the discrete particles are replaced be a continuum density, we see that the energy functional (\ref{1.4}) results.

We remark that (\ref{A3}) with the change of variables $y_l = x_l^{1/M}$ is an example of the
so-called Laguerre Muttalib--Borodin ensemble \cite{Mu95,Bo98}. Its relationship to random
matrix products has shown itself from other viewpoints in the recent works
\cite{KS14, FW14}.

\section*{Acknowlegements} The work of PJF was supported by the Australian Research Council, grant
DP140103104. The work of DZL  was  supported by the National Natural Science Foundation of China, grants  11301499 and  11171005. The work of PZJ was supported by the European Research Council, grant 278124, and the Australian Research Council, grant DP140102201.


\providecommand{\bysame}{\leavevmode\hbox to3em{\hrulefill}\thinspace}
\providecommand{\MR}{\relax\ifhmode\unskip\space\fi MR }
\providecommand{\MRhref}[2]{%
  \href{http://www.ams.org/mathscinet-getitem?mr=#1}{#2}
}
\providecommand{\href}[2]{#2}

\end{document}